\documentclass[10pt]{article}
\usepackage[utf8]{inputenc}
\usepackage{newunicodechar,graphicx}
\DeclareRobustCommand{\okina}{%
  \raisebox{\dimexpr\fontcharht\font`A-\height}{%
    \scalebox{0.8}{`}%
  }%
}
\newunicodechar{ʻ}{\okina}
\usepackage{enumitem}
\setlist[itemize,enumerate]{noitemsep, topsep=0pt, leftmargin=1.0em}

\usepackage[letterpaper]{geometry}
\usepackage{hicss}
\usepackage{times}
\usepackage[none]{hyphenat}
\usepackage{url}
\usepackage{latexsym}
\usepackage{indentfirst}
\usepackage{graphicx}
\usepackage{soul}
\usepackage{booktabs}
\usepackage{multirow}
\usepackage{float}
\usepackage{tcolorbox}
\usepackage{tabularx} 
\usepackage{balance}
\usepackage{caption}
\usepackage{booktabs}
\usepackage[hidelinks,bookmarks=true]{hyperref}
\usepackage[style=apa]{biblatex}
\usepackage{background}
\backgroundsetup{
  position=current page.east,
  angle=-90,
  nodeanchor=east,
  vshift=-4mm,
  opacity=0.5,
  scale=3,
  contents=PPREPRINT
}

\title{The Impact of Generative AI-Powered Code Generation Tools on Software Engineer Hiring: Recruiters' Experiences, Perceptions, and Strategies}

 \author{Alyssia Chen  \\
  University of Hawaiʻi at Mānoa \\
  {\underline{\href{mailto:abc8@hawaii.edu}{abc8@hawaii.edu}} } \\ \\
  Dan Port  \\
  University of Hawaiʻi at Mānoa \\
  {\underline{\href{mailto:peruma@hawaii.edu}{dport@hawaii.edu}} } \\  \And
  Timothy Huo  \\
  University of Hawaiʻi at Mānoa \\
  {\underline{\href{mailto:thuo@hawaii.edu}{thuo@hawaii.edu}} } \\ \\
  Anthony Peruma \\
  University of Hawaiʻi at Mānoa \\ 
  {\underline{\href{mailto:dport@hawaii.edu}{peruma@hawaii.edu}} }\\ \And
  Yunhee Nam  \\
  University of Hawaiʻi at Mānoa \\
  {\underline{\href{mailto:yunheen@hawaii.edu}{yunheen@hawaii.edu}} } \\}
  
\date{}

\newcommand{\RQA}{\textbf{RQ1}: How familiar are recruiters with GenAI code generation tools?}

\newcommand{\RQB}{\textbf{RQ2}: How are organizations and recruiters adapting their candidate screening and skills evaluation processes in response to GenAI code generation tools?}

\newcommand{\RQC}{\textbf{RQ3}: To what extent is a candidate's experience in using GenAI code generation tools helpful in evaluating them?}

\newcommand{\RQD}{\textbf{RQ4}: What is the perception within the software engineering industry on the importance of integrating GenAI code generation tools into the computer science (or related) curricula?}

\begin{document}
\maketitle
\begin{abstract}
The rapid advancements in Generative AI (GenAI) tools, such as ChatGPT and GitHub Copilot, are transforming software engineering by automating code generation tasks. While these tools improve developer productivity, they also present challenges for organizations and hiring professionals in evaluating software engineering candidates' true abilities and potential. Although there is existing research on these tools in both industry and academia, there is a lack of research on how these tools specifically affect the hiring process. Therefore, this study aims to explore recruiters' experiences and perceptions regarding GenAI-powered code generation tools, as well as their challenges and strategies for evaluating candidates. Findings from our survey of 32 industry professionals indicate that although most participants are familiar with such tools, the majority of organizations have not adjusted their candidate evaluation methods to account for candidates' use/knowledge of these tools. There are mixed opinions on whether candidates should be allowed to use these tools during interviews, with many participants valuing candidates who can effectively demonstrate their skills in using these tools. Additionally, most participants believe that it is important to incorporate GenAI-powered code generation tools into computer science curricula and mention the key risks and benefits of doing so.
\end{abstract}

\subsubsection*{Keywords:}

practitioner survey, code generation, recruiter, hiring in computing, generative AI

\section{Introduction}
\label{sec:introduction}
Recent advancements in Artificial Intelligence (AI) research have resulted in significant progress in Generative AI (GenAI) tools and techniques. These developments are significantly impacting various industries by automating processes that previously required human intervention, thereby streamlining operations and improving efficiencies (\cite{mckinsey2024}). One such industry where the impact of GenAI is particularly noteworthy is the software engineering industry. With their ability to generate code and provide intelligent auto-completion, GenAI tools, such as ChatGPT\footnote{\url{https://chat.openai.com}} and GitHub Copilot\footnote{\url{https://github.com/features/copilot}} are revolutionizing the software industry by changing the way developers write code (\cite{Ebert2023}). These GenAI-powered code generation tools enhance developer productivity by automating repetitive tasks, identifying bugs or poor-quality code, and providing solutions (\cite{Sauvola2024,Liang2024}). However, there are also drawbacks to consider, such as hallucinations, ethical and legal concerns related to using GenAI tools to generate code, potential biases in the generated code, and the risk of overreliance, which could lead to a decline in coding skills and a lack of understanding of underlying concepts (\cite{Bull2024,Russo2024,Fan2023}). 

As these GenAI-powered code generation tools become more prevalent and widely accessible, it is crucial to understand how they influence the skills and qualifications required for software engineers and how organizations and hiring professionals have had to adapt their evaluation approaches. For instance, evaluating a candidate's programming, problem-solving, innovation, and creativity skills becomes challenging if they rely heavily on GenAI-powered code generation tools.

\subsection{Goal \& Research Questions}
\label{sec:goal}
This shift towards AI-augmented software development (\cite{Ozkaya2023}) necessitates a reevaluation of traditional evaluation techniques to ensure that organizations are able to accurately assess a software engineering candidate's true abilities and potential. However, given the recent emergence of these tools, there is a lack of research on how the hiring process should evolve to assess candidates who have experience utilizing such tools. Therefore the goal of this study is to \textit{investigate how organizations and hiring professionals have adapted their hiring strategies and policies to account for GenAI-powered code generation tools, and to understand their perceptions of such tools}. Hence, we answer the following research questions (RQs): 

\noindent\textbf{\RQA} This RQ aims to understand recruiters' awareness of the latest advancements in AI code generation tools and ascertain if they are prepared to evaluate candidates who utilize these tools.

\noindent\textbf{\RQB} As AI code generation tools become more widespread, this RQ explores how organizations and recruiters are adapting their evaluation strategies to ensure fair and accurate assessments and the factors they consider when making hiring decisions.

\noindent\textbf{\RQC} This RQ investigates the value recruiters place on a candidate's knowledge and experience with AI code generation tools to gain insights into changing skill requirements and the importance of proficiency with these tools in the hiring process.

\noindent\textbf{\RQD} As AI code generation tools become more prevalent in the software engineering industry, this RQ aims to gather insights on how academia should adapt to ensure students are prepared for the evolving demands of the software engineering industry.

\subsection{Contribution}
To answer our RQs, we surveyed 32 industry professionals who are involved in recruiting software engineers to understand how they evaluate candidates in the era of GenAI-powered code generation tools. We envision our findings providing valuable insights into the changing landscape of software engineer recruitment and offering practical recommendations for employers, candidates, and academia to navigate the challenges and opportunities presented by such GenAI-powered tools.

\section{Related Work} 
\label{sec:related}
Existing research has studied the software engineering industry's hiring process, including candidate skills, resume evaluation, and the hiring process from the student perspective. However, limited research exists on how AI code generation tools have impacted the hiring process and candidate evaluation.

\textcite{Singh2021} investigated whether computer-science-oriented, engineering graduate students' knowledge in reliability and safety engineering were at the level industry professionals expected. To do so, both students and practitioners were interviewed in the study, and their analysis revealed that most of the graduate students' knowledge was not up to expectations. \textcite{Petersheim2023} conducted a study with both recruiters and computer science students, and both groups were given entry-level computer science resumes to review. It was found that recruiters and students prioritized certain resume items differently and spent varying amounts of time in sections of the resume. \textcite{garousi2019aligning} conducted a meta-analysis of 35 studies to identify the most important skills needed by software engineering graduates and found ``requirements, design and testing'' to be the most critical and found ``configuration management, models and process'' to be the knowledge graduates did not know enough about.

\textcite{Chinn2008} examined students' responses to an in-class assignment requiring them to evaluate and hire fictitious candidates. This allowed the authors to identify the hiring criteria that students prioritized, which included technical skills, soft skills, personal traits, previous employment, and career considerations. \textcite{Lunn2024} focused on the whole hiring process itself from a student's point of view and discovered how students prepare and get support, with emphasis on the inclusivity of the process. \textcite{adnin2022hiring} examined the hiring process from an employer's perspective, and through interviews, they discovered sources of bias and significant preferences for strong technical skills. 
\textcite{Stepanova2022} analyzed the survey responses of more than 200 recruiters in the US to learn more about the hiring process and recruiters' hiring choices. \textcite{alekseeva2021demand} found a significant increase in the demand for AI skills in most industries and occupations in the U.S. labor market from 2010 to 2019. A practitioner survey by \textcite{Akdur2022}  identified the most in-demand skills for software engineering. \textcite{scaffidi2018employers} interviewed employers to understand the soft skills and personal attributes new graduates need to impress recruiters.

\section{Method}
\label{sec:method}
This section presents our survey design, participant recruitment, and response analysis methodology. As our study involves working with human subjects, we obtained approval from our institution's Institutional Review Board of the Office of Research Compliance.

\subsection{Survey Design}
We utilized Qualtrics to design and host the survey and set it up to permit only one response per participant. Our survey comprised 24 questions aimed at collecting data on participants' demographics, their experiences with and perceptions of AI code generation tools for candidate evaluation, and their perspectives on integrating such tools into academic curricula. These questions were developed based on the objectives of our study, as discussed in Section \ref{sec:goal}, and our review of relevant literature, as discussed in Section \ref{sec:related}. Finally, as best practice, we conducted a pilot run with a couple practitioners before publishing the survey to identify potential issues with the questionnaire (\cite{linaaker2015guidelines}). Based on their feedback, we reworded some questions to improve clarity and re-ordered a few. Table \ref{Table:SurveyQuestions} shows the published survey questions, the question type, whether a response is required, and any logic/notes applicable to each question. The complete questionnaire and survey responses are included in our artifact package (\cite{ProjectArtifacts}).

\begin{table*}
\centering
\caption[xxx]{Below are the questions that are part of the survey. The questionnaire including the answer options for the single-choice and multi-choice questions, is available in our artifact package \cite{ProjectArtifacts}.}
\vspace{-2mm}
\label{Table:SurveyQuestions}
\resizebox{\textwidth}{!}{%
\begin{tabular}{lp{0.7\linewidth}llp{0.2\linewidth}}
\toprule
\multicolumn{1}{c}{\textbf{No.}} &
  \multicolumn{1}{c}{\textbf{Question}} &
  \multicolumn{1}{c}{\textbf{Type}} &
  \multicolumn{1}{c}{\textbf{Required}} &
  \multicolumn{1}{c}{\textbf{Notes}} \\ \midrule
1 &
  What is your gender? &
  Single-Choice &
  Yes &
  Includes “Other” free-text option \\ \midrule
2 &
  What is your age? &
  Single-Choice &
  Yes &
   \\ \midrule
3 &
  How many years of industry experience do you have? &
  Single-Choice &
  Yes &
   \\ \midrule
4 &
  Which job category closely matches your current position? &
  Single-Choice &
  Yes &
  Includes ``Other'' free-text option \\ \midrule
5 &
  Where are you employed? &
  Single-Choice &
  Yes &
  Includes ``Other'' free-text option \\ \midrule
6 &
  How many employees work in your establishment? &
  Single-Choice &
  Yes &
   \\ \midrule
7 &
  Which of the following industries most closely matches the one in which you are employed? &
  Single-Choice &
  Yes &
  Includes ``Other'' free-text option \\ \midrule
8 &
  What is your role in the hiring process? &
  Free Text &
  Yes &
   \\ \midrule
9 &
  Which of these AI code generation tools are you familiar with? &
  Multi-Choice &
  Yes &
  Includes ``Other'' free-text option \\ \midrule
10 &
  How frequently do you use an AI code generation tool for a work project or task? &
  Single-Choice &
  Yes &
  Show only if anything but ``None'' in Q9 was selected
   \\ \midrule
11 &
  With the increasing prevalence of AI code-generation tools, has your organization developed official guidelines or policies related to these specific tools when evaluating candidates? &
  Single-Choice &
  Yes &
   \\ \midrule
12 &
  If possible, can you provide a summary or key points of the guidelines/policies to evaluate candidates with regard to AI code generation tools? &
  Free Text &
  No &
  Shown only if ``Yes'' is selected in Q11 \\ \midrule
13 &
  Have you changed the criteria you utilize to evaluate a candidate's coding skills due to the wide availability of AI code-generation skills? &
  Single-Choice &
  Yes &
  Shown only if ``No'' or ``Not Sure'' is selected in Q11 \\ \midrule
14 &
  If possible, can you provide a summary or key points of your candidate code evaluation criteria? &
  Free Text &
  No &
  Shown only if ``Yes'' is selected in Q13 \\ \midrule
15 &
  Have you permitted candidates to utilize AI code generation tools in a technical interview? &
  Single-Choice &
  Yes &
   \\ \midrule
16 &
  Should candidates be allowed to utilize AI code generation tools during technical interviews? &
  Single-Choice &
  Yes &
   \\ \midrule
17 &
  Please let us know why you selected this answer option. &
  Free Text &
  Yes &
  Shown only if ``Yes'' or ``No'' is selected in Q16 \\ \midrule
18 &
  To what extent do you believe AI code generation tools pose challenges in accurately assessing a candidate's coding proficiency, analytical skills, and overall experience? &
  Single-Choice &
  Yes &
   \\ \midrule
19 &
  How frequently do you directly ask a candidate about their personal experience and skills in using AI code generation tools? &
  Single-Choice &
  Yes &
   \\ \midrule
20 &
  How much of a preference do you give to candidates who demonstrate skills in using AI code generation tools? &
  Single-Choice &
  Yes &
   \\ \midrule
21 &
  From your perspective, what critical skills should candidates possess to utilize AI code generation tools effectively? &
  Free Text &
  Yes &
   \\ \midrule
22 &
  How important is it for universities to integrate education on AI code generation tools into their curricula? &
  Single-Choice &
  Yes &
   \\ \midrule
23 &
  What, if any, benefits do you foresee in integrating AI code generation tools into academic courses? &
  Free Text &
  No &
   \\ \midrule
24 &
  What, if any, risks or concerns do you foresee in integrating AI code generation tools into academic courses? &
  Free Text &
  No &
   \\ \bottomrule
\end{tabular}%
}
\end{table*}

\subsection{Survey Participants}
To ensure the reliability of the responses, we made sure that survey participants were involved in the recruitment process for software engineers or similar positions. To accomplish this, we contacted representatives of companies that took part in the Spring 2024 Career Fair hosted by the Information and Computer Sciences Department at the University of Hawaiʻi at Mānoa. The fair had 30 organizations from various sectors, with an average of two employees from each. Around 174 computer science students attended, looking for technology sector jobs.

To recruit participants, the authors met with industry representatives at the Career Fair. After confirming that they were involved in the hiring of software engineers, the representatives were invited to participate in the survey. They were provided with a QR code to access the online survey, or offered a laptop or tablet to complete the survey. Additionally, some participants also referred colleagues and friends, who were then emailed the survey link. This approach of convenience and snowball sampling enabled us to identify and invite individuals with the necessary qualifications required for our study. Before starting the survey, participants had to agree to an informed consent document. Participation was voluntary, and no compensation was offered.

\subsection{Data Analysis}
We analyzed the survey data using both quantitative and qualitative methods (\cite{Wagner2020}). Our quantitative analysis involved standard statistical techniques, while for qualitative analysis, we reviewed participants' open-ended responses to identify common themes. To ensure reliability, three authors independently reviewed and categorized the responses, discussed any discrepancies, and reached a consensus. We elaborate on the specific techniques when answering our RQs.

\section{Results}
\label{sec:results}
This section presents our RQ results. We first report the number of responses received and participant demographics before answering our RQs. It should be noted that the responses received from participants in the pilot run are not in the RQ results.

\subsection*{\textbf{Survey Responses}}
The survey was available to the public from February 2024 to May 2024, and we gathered 39 responses within this period. However, not all participants responded to all the questions. To maintain consistency in our analysis, we only included participants who answered all the required questions, leading to 32 valid responses. Hence, our results focus solely on these 32 responses.

\subsection*{\textbf{Participant Demographics}}
We divide our reporting of participants' demographics (survey questions \#1 to \#8) into two parts: general demographics and work experience. 

\textit{\textbf{General Demographic:}} Among the 32 participants, 17 (53.13\%) identified as male, 14 (43.75\%) as female, and 1 (3.13 \%) preferred not to disclose. In terms of age, 2 (6.25\%) were between the age range of 18-24, 12 (37.5\%) between 25-34, 9 (28.13\%) between 35-44, 3 (9.38\%) between 45-54, and 6 (18.75\%) between 55-64. 

\textit{\textbf{Work Experience:}} For years of industry experience, almost half (46.88\%) of participants reported more than 15 years, 2 (6.25\%) report 11-15 years, 6 (18.75\%) report 6-10 years, 4 (12.5\%) report 3-5 years, and 5 (15.63\%) report 1-2 years. When asked which job category they closely match with, 13 (40.63\%) reported ``HR/Talent Acquisition/Hiring Manager/ Or Similar'', 5 reported ``Software Engineering/Software Architect/ Or Similar'', 7 reported ``Team Lead/Project Manager/Account Manager/ Or Similar'', and another 7 reported ``Other''. Those who report ``Other'' listed other specific technical and communication roles such as Cybersecurity, Communications \& Marketing, and Health Information Technology. Lastly, 12 (37.5\%) participants are employed in the private-for-profit sector, 2 (6.25\%) in the private-not-for-profit sector, 1 (3.13\%) in local government, 11 (34.38\%) in state government, and 6 (18.75\%) in federal government. 

The demographic details highlight that, despite the relatively small sample size of 32 participants, they represent various industries and possess extensive experience within their respective fields. As such, their feedback offers valuable insights for our study.

\subsection*{\RQA}

For this RQ, we analyze responses from survey questions \#9 and \#10 regarding their familiarity and usage of AI code generation tools.

\begin{table}[htbp]
\centering
\caption{Familiarity with AI code generation tools.}
\vspace{-2mm}
\label{Table:AIFamiliarity}
\resizebox{\columnwidth}{!}{%
\begin{tabular}{@{}lrr@{}}
\toprule
\multicolumn{1}{c}{\textbf{Answer Options}} & \multicolumn{1}{c}{\textbf{Count}} & \multicolumn{1}{c}{\textbf{Percentage}} \\ \midrule
ChatGPT
& 28 & 60.87\% \\
Github Copilot
& 8 & 17.39\% \\
Amazon CodeWhisperer 
& 2  & 4.35\%  \\
Replit AI
& 2  & 4.35\%  \\ 
Other
& 2  & 4.35\% \\ \bottomrule
\end{tabular}%
}
\end{table}

Survey question \#9 is a multi-choice question with an ``Other'' free text option, aiming to gauge recruiters' familiarity with AI code generation tools. As shown in Table \ref{Table:AIFamiliarity}, the AI code generation tools participants were familiar with include ``ChatGPT'' (28 or 60.87\%), ``Github Copilot'' (8 or 17.39\%), ``Amazon CodeWhisperer'' (2 or 4.35\%), ``Replit AI'' (2 or 4.35\%), and ``Other'' (2 or 4.35\%). The chocies under ``Other'' were ``Bard (Gemini)'' and ``Pilot AI''. Additionally, 4 (8.70\%) participants reported they were not familiar with AI code generation tools. As this is a multi-choice question, respondents had an average of 1.31 answer options chosen, with the most common combination being ``ChatGPT'' and ``Github Copilot'' selected eight times. The next most frequent combinations were ``ChatGPT'' and ``Amazon CodeWhisperer'', ``ChatGPT'' and ``Replit AI'', and ``Github Copilot'' and ``Replit AI'', each occurring twice. 

\begin{figure}[]
 	\centering
 	\includegraphics[trim=0cm 0cm 0cm 0cm,clip,scale=0.39]{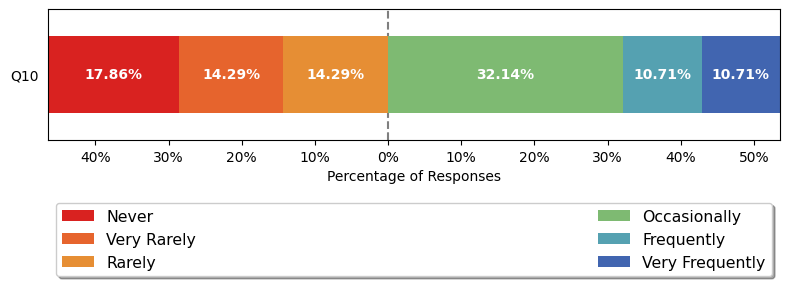}
 	\caption{Frequency of AI code generation tool usage for work}
 	\label{Figure:q12}
\end{figure}

Survey question \#10 is a Likert (i.e., single-choice) question examining the frequency with which recruiters use AI code generation tools for work projects and tasks. As shown in Figure \ref{Figure:q12}, more than half of the participants reported using AI code generation tools, with 3 (10.71\%) selecting ``Very Frequently'', another 3 (10.71\%) selecting "Frequently", and 9 (32.14\%) selecting ``Occasionally''. In contrast, 4 (14.28\%) report ``Rarely'', another 4 (14.28\%) report ''Very Rarely``, and 5 (17.86\%) report ``Never''. 

\begin{tcolorbox}[top=0.5pt,bottom=0.5pt,left=1pt,right=1pt]
\textbf{Summary for RQ1.}
ChatGPT stood out as the most familiar AI code generation tool, with recognition at 60.87\%, but recognition significantly dropped for other tools, followed by Github Copilot at 17.39\%. Moreover, a majority (53.57\%) of respondents indicated occasional to very frequent usage of AI code generation tools for work. Conversely, 28.57\% reported rarely or very rarely using such tools, while another 17.86\% reported never using them.
\end{tcolorbox}

\subsection*{\RQB}
This RQ has two parts. The first examines how organizations approach AI code generation tools, while the second examines the evaluator's experience and perception of such tools when evaluating candidates.   

\vspace{2mm}
\noindent\textbf{Organizational Policy:}

Through survey questions \#11 and \#12, we gain insight into how organizations have adapted to these tools when evaluating candidates. Survey question \#11 is a single-choice question designed to determine if organizations are adjusting to the emergence of AI code generation tools when hiring software engineers. We found that 5 (15.63\%) participants selected ``Yes'', 21(65.63\%) selected ``No'', and 6(18.75\%) selected ``Not Sure''. Participants who selected ``Yes'' were then asked to provide a summary of their organization's guidelines/policies regarding the use of AI code generation tools in candidate evaluation. Only one participant responded, cautioning against the use of AI.

\vspace{2mm}
\noindent\textbf{Recruiter Perception \& Experience:}

Next, we examine how individual recruiters evaluate candidates regarding AI code generation tools. First, from survey question \#13, 4 participants (14.81\%) reported that they have changed the criteria they use to evaluate a candidate’s coding skills, 17 (62.96\%) reported that they have not, and 6 (22.22\%) selected 'Not applicable.'. Participants who said they had changed their evaluation criteria were prompted with an optional survey question, \#14, and asked, if possible, to provide a summary of their evaluation criteria. There were only three responses to this question, which we reviewed and identified the following two main categories: 
\begin{itemize}
    \item \textbf{Standardized Evaluation Criteria}: One participant mentioned they follow a set of predefined and uniform standards to evaluate candidate performance (e.g., \textit{``Federal government uses KSAs''}).
    \item \textbf{Verbal Discussion}: Two participants mention they rely on verbal techniques to evaluate candidates' understanding of coding and problem-solving skills (e.g., \textit{``We rely more on verbal interviews to discuss methodology and general approaches to tasks and problems rather than coding or written exercises''}).
\end{itemize}

Moving on, survey question \#15 asks participants whether they allow candidates to utilize AI code generation tools in technical interviews. Only 1 (3.13\%) participant selected "Yes", while 20 (62.50\%) selected "No", and 11 (34.38\%) selected "Not applicable". Next, in survey question \#16, we ask participants if candidates should be allowed to utilize AI code generation tools during technical interviews. The results show that 5 participants (25\%) report ``Yes'', 7 participants (35\%) report ``No'', and 8 participants (40\%) report ``Not Sure''.

In survey question \#17, participants who reported ``Yes'' and ``No'' in question \#16 were asked to explain their choices.

Examining the free text responses for the ``Yes'' option, we identified the below three categories:
\begin{itemize}
    \item \textbf{AI is a tool}: Three participants see AI code generation tools as another resource for software engineers. They believe candidates should be permitted to use these tools in interviews as they are widely available and prevalent. (e.g., \textit{``AI is another tool that can be used by the applicant''}).
    \item \textbf{Promotes Fairness}: One participant mentioned 
    that allowing AI code generation tools in technical interviews is a fair approach and unbiased. 
    \item \textbf{Demonstrates Proficiency of Current Technologies}: One participant mentions AI enables candidates to showcase their familiarity with the most recent technologies during the interview (e.g., \textit{It allows candidates to display their knowledge of the most current technologies out there}).
\end{itemize}

For responses associated with ``No'', we identified the following three main categories: 
\begin{itemize}
    \item \textbf{Challenge Evaluating Candidates True Technical Skills}: Four participants mention that they want to evaluate a candidate's knowledge and skill without the assistance of AI tools. These tools hinder an evaluator's ability to determine if candidates possess the necessary expertise to perform their roles independently (e.g., \textit{``Certain roles need people to know the information off the top of their head without the use of AI''}).
    \item \textbf{Concerns Over Quality of AI-Generated Code}: One participant expressed doubts about the accuracy and reliability of AI-generated code. (e.g., \textit{``Because being able to code without AI is a skill necessary to determine AI hallucinations''})
    \item \textbf{Challenge Evaluating Candidates Cognitive Skills}: Two participants highlighted the importance of assessing candidates' critical thinking and problem-solving skills during technical interviews. AI code generation tools may hinder the evaluation of these crucial abilities (e.g., \textit{``Our technical interviews isn't about writing a lot of code, its about your thought process to solve the problem''}).
\end{itemize}

\begin{figure}[]
 	\centering
 	\includegraphics[trim=0cm 0cm 0cm 0cm,clip,scale=0.39]{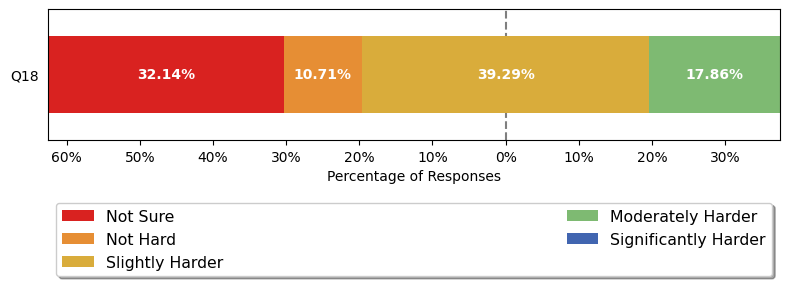}\vspace{-2mm}
 	\caption{Perceived challenges in assessing candidate skills with AI code generation tools}
 	\label{Figure:q24}
\end{figure}
Lastly, survey question \#18 examines the challenge AI code generation tools have in assessing a candidate's coding proficiency, analytical skills, and overall experience. According to Figure \ref{Figure:q24}, excluding 4 participants who chose "Not Applicable", half of the participants report these tools will pose more of a challenge, with 5 (17.86\%) selecting ``Moderately Harder'' and 11 (39.29\%) selecting ``Slightly Harder''. In contrast, only 3 (10.71\%) selected ``Not Hard'' and 9 (32.14\%) report ``Not Sure''.

\begin{tcolorbox}[top=0.5pt,bottom=0.5pt,left=1pt,right=1pt]
\textbf{Summary for RQ2.}
Most organizations and recruiters have not adapted their candidate evaluation techniques to account for AI code generation tools. For those that have, one common means to evaluate candidates is a verbal discussion about their code. Opinions on allowing candidates to use AI code generation tools during interviews were divided. Those in favor highlighted AI tools as additional resources, promoting fairness and showcasing candidates' familiarity with modern technology. Those against emphasized the importance of assessing candidates' genuine skills without AI assistance and prioritizing problem-solving approaches over extensive code writing during interviews. Lastly, participants responded that assessing candidates' skills with AI code generation tools as moderately or slightly more challenging, with only a minority expressing confidence.
\end{tcolorbox}

\subsection*{\RQC}
To answer this RQ, we analyzed participants' responses to question \#19 to determine how frequently they inquire about candidates' experience and skills in AI code generation tools. Subsequently, with survey question \#20, we investigated whether any preference is shown towards candidates demonstrating expertise in AI code generation tools. Lastly, in survey question \#21, we explored their opinions on the critical skills that candidates should possess to effectively utilize AI code generation tools.

\begin{figure}[]
 	\centering
 	\includegraphics[trim=0cm 0cm 0cm 0cm,clip,scale=0.39]{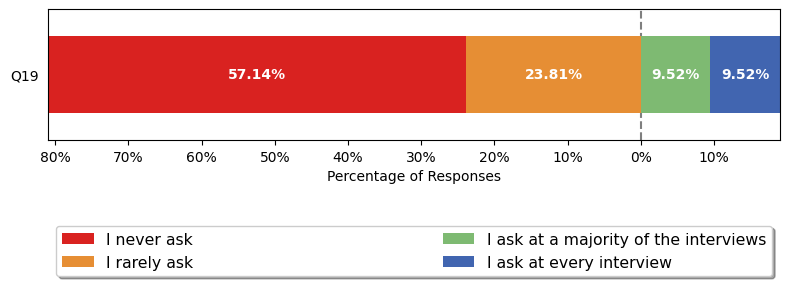}\vspace{-2mm}
 	\caption{Frequency of directly asking a candidate about their personal experience and skills in using AI code generation tools}
 	\label{Figure:q23}
\end{figure}

\begin{figure}[]
 	\centering
 	\includegraphics[trim=0cm 0cm 0cm 0cm,clip,scale=0.39]{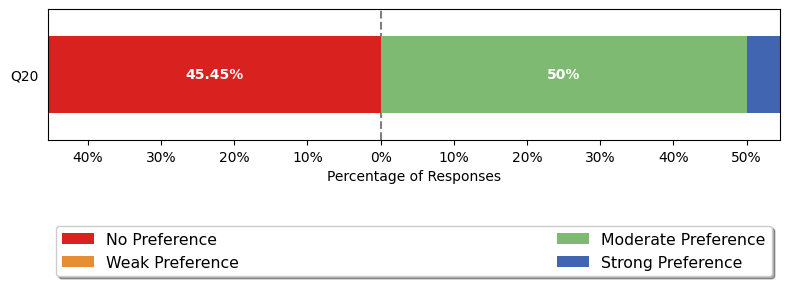}\vspace{-2mm}
 	\caption{Extent of preference given to candidates who demonstrate skills in using AI code generation tools}
 	\label{Figure:q22}
\end{figure}

As shown in Figure \ref{Figure:q23}, excluding the 11 participants who wrote ``Not Applicable'', 17 (53.13\%) of the participants never or rarely asked candidates about their experience in AI code generation tools, and 2 (6.25\%) asked at every interview. In Figure \ref{Figure:q22}, excluding the 10 who answered ``Not Applicable,'' more than 50\% have a moderate or strong preference for candidates who demonstrate AI code generation skills.

\begin{figure}[]
 	\centering
 	\includegraphics[trim=0cm 0cm 0cm 0cm,clip,scale=0.39]{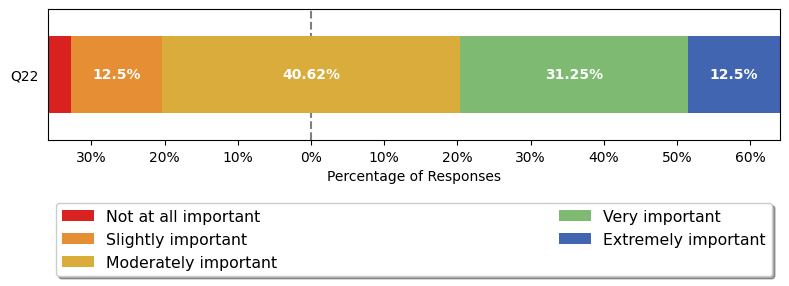}\vspace{-2mm}
 	\caption{Importance of integrating AI code generation tools into computer science curricula.}
 	\label{Figure:q26}
\end{figure}

Next, when asked to describe the critical skills candidates should have to effectively use AI code generation tools (survey question \#21), we identified five categories:
\begin{itemize}
    \item \textbf{Critical Thinking \& Problem-Solving Skills}: Five participants mentioned the importance of candidates being able to analyze problems, understand fundamental concepts, and general problem-solving skills (e.g., \textit{``Critical Thinking and Problem Solving Skills''}).
    
    \item \textbf{Prompt Engineering}: Five participants highlighted that candidates should be able to ask the right questions and provide clear, context-rich prompts to the AI tool to obtain the most relevant and accurate results (e.g., \textit{``The ability to properly prompt and provide accurate and detailed context to the tool to get the best possible outcomes''}).
    
    \item \textbf{Evaluate AI Response}: Six participants reported that knowing the limitations and shortcomings of responses from AI code generation tools is an essential skill that candidates should possess (e.g., \textit{``They need to be able to tell when the generated code is not just correct technically but achieves the requirements of the software''}).
     
    \item \textbf{Familiarity of AI Tools}: Two participants mentioned the importance of knowing how to effectively use the tools/technologies as an important skill for candidates (e.g., \textit{``Should be able to navigate platforms and provide demonstration to audiences''}).
    
    \item \textbf{Fundamental Computer Science Knowledge}: One participant mentioned the importance of core computer science courses (e.g., \textit{``applied statistics, data structures''}).
\end{itemize}

\noindent Two participants did not answer the question, and 11 stated that they were unsure about the required skills.

\begin{tcolorbox}[top=0.5pt,bottom=0.5pt,left=1pt,right=1pt]
\textbf{Summary for RQ3.}
Most participants never inquire about candidates' AI code generation tool experience during interviews. However, around half of the participants give candidates a moderate to strong preference if they demonstrate skills in using these tools. Key skills deemed necessary for effective utilization of these tools include critical thinking and problem-solving, prompt engineering, evaluating AI outputs, familiarity with the tools, and fundamental computer science knowledge. Additionally, some participants were uncertain about the necessary skills.
\end{tcolorbox}

\subsection*{\RQD}
In this RQ, participants were asked about the importance of integrating AI code generation tools into computer science curricula (question \#22) and to outline the benefits (question \#23) and risks (question \#24) they foresee in doing so.

In terms of importance, as shown in Figure \ref{Figure:q26}, only a small percentage believe it is not at all important to have educational institutes integrate AI code generation tools into the course curricula, and almost 45\% believe it is very important or extremely important.

We next report on the responses participants provided for the benefits and risks survey question. Three of the authors reviewed the 32 free-text responses to identify common themes. Below we describe the categories our analysis yields:

\noindent\textbf{Benefits:}
\begin{itemize}
    \item \textbf{Improved Productivity}: Five participants mentioned that AI code generation tools can help students and educators save time on repetitive tasks, such as writing boilerplate code and grading. Instead, they can have more time to focus on problem-solving skills or other high-level concepts (e.g., \textit{``More time to focus on unique problem sets and overall quality''}). 
   
    \item \textbf{Equitable Access}: Four participants touched on that AI code generation tools have the potential to support students who may not have had prior exposure to programming or those who are transitioning from other fields (e.g., \textit{``Easier time obtaining knowledge before going into the field''}). 

    \item \textbf{General Awareness \& Preparedness}: Ten participants mentioned that incorporating AI code generation tools into courses can potentially help students stay ahead of the curve, better prepare for the future of work, and enhance their understanding of how AI can be utilized in various contexts (e.g., \textit{``AI is inevitable, if anything, it will be developed so it should be incorporated into a academic courses''}).

    \item \textbf{Industry Relevance and Preparedness}: Three participants specifically mentioned how exposure to AI code generation tools in academia will better prepare students for the real world, where they will encounter such tools (e.g., \textit{``College should be teaching students real world applications. Using AI code generation tools is a real world use case''}).
\end{itemize}

\noindent Ten participants either did not provide a response or stated that they were unsure about the benefits. 

\noindent\textbf{Risks:}
\begin{itemize}
    \item \textbf{Academic Dishonesty}: Six participants had concerns about academic integrity, including instances of cheating and plagiarism due to the use of these tools (e.g., \textit{``Potential for cheating and plagiarism''}).

    \item \textbf{Challenges for Educators}: Four participants mentioned that these tools will negatively impact educators as they would need to spend time restructuring their courses and face difficulty in evaluating students' skills (e.g., \textit{``Basically professors are going to have a harder time determining what students actually learned''}).    

    \item \textbf{Student Overreliance}: Six participants highlight the risk of students becoming overly reliant on these tools, potentially lacking essential computing and problem-solving skills. (e.g., \textit{``Dependency on AI for all tasks resulting in lack of problem solving skills''}).   

    \item \textbf{Lack of Responsible Use}: Four participants were concerned about the potential misuse of these tools (e.g., \textit{``Responsible use and malicious code''}).
\end{itemize}

\noindent Three participants indicated that there are no risks, while nine 
either did not provide a response or stated that they were unsure about the risks. 

\begin{tcolorbox}[top=0.5pt,bottom=0.5pt,left=1pt,right=1pt]
\textbf{Summary for RQ4.}
Most participants are of the opinion that integrating education on AI code generation tools into computer science curricula is important. The benefits identified include improved productivity, equitable access, general awareness and preparedness, as well as industry relevance and preparedness. However, there are also perceived risks such as academic dishonesty, challenges for educators, student overreliance, and lack of responsible use. Some participants were unsure or did not provide a response regarding both the benefits and risks.
\end{tcolorbox}

\section{Discussion \& Takeaways}
\label{sec:discussion}
With the recent emergence of GenAI-based code generation tools and the limited research on their impact on hiring software engineers, our exploratory study offers valuable insights that can benefit various stakeholders, including practitioners, educators, researchers, and students.

Our RQ1 findings indicate that most practitioners are familiar with and have experience in using these tools, especially ChatGPT. However, the relative newness of these technologies presents a challenge for organizations and recruiters in effectively evaluating candidates' true skills and abilities, considering the potential influence of these tools, as revealed by our RQ2 results. Further, the lack of industry standards and guidelines on assessing candidates' actual capabilities in light of the increasing use of AI code generation tools makes it harder for recruiters to make informed decisions. Moving on, insights from RQ3 highlight the importance practitioners place on a candidate's cognitive skills, including critically analyzing the tool's output (\cite{Wang2024}). Finally, based on RQ4, it is encouraging to see that practitioners, educators, and students share similar views on the risks and benefits of these tools (\cite{Zastudil2023, Sheard2024}).

\noindent\textit{\textbf{Takeaways}}
\begin{itemize}
    \item There is a need for \textit{close collaboration between industry and academia to develop curricula that effectively prepare students for the AI-augmented future of software engineering}. This collaboration should focus on designing learning experiences that go beyond simply introducing and teaching students about GenAI tools. It's important for students to understand the challenges of GenAI tools and work on assignments resembling real-world software engineering scenarios to develop the skills needed to use GenAI tools effectively.
    \item It is crucial for \textit{organizations to continually update their candidate hiring and evaluation techniques} to effectively gauge candidates' skills and abilities in the current AI-augmented era. This includes training their hiring team to understand the capabilities, limitations, and potential impact of GenAI tools on the software development process. Organizations should also focus on evaluating candidates' cognitive abilities, such as their ability to engage in chain-of-thought reasoning, prompt engineering, and code review skills, among others.
\end{itemize}

\section{Threats To Validity}
\label{sec:threats}
The generalizability of the findings may be affected by factors such as the number of participants, geographical location, and self-selection bias. However, as per the demographic data, these participants come from various industries and have been verified to be involved in hiring software engineers. Additionally, the anonymous nature of our survey encouraged participants to provide honest and unbiased responses. Although the study includes open-ended questions, the depth of qualitative data may be limited. Conducting in-depth interviews or focus groups could provide more comprehensive insights. Moreover, there is a possibility that participants may have interpreted the survey questions differently, resulting in variations in responses. Lastly, given the rapidly evolving nature of the AI field, the survey responses offer valuable insights at this specific point in time, serving as a platform for future studies to examine the evolving hiring practices.

\section{Conclusion \& Future Work}
\label{sec:conclusion}
This study provides early insights into the extent to which organizations and recruiters are adapting their evaluation techniques to assess the true abilities of candidates and their perceptions of these tools. Furthermore, our findings show a consensus among industry professionals on the importance of integrating AI code generation tools into university curricula, recognizing the need to prepare students for the AI-augmented future of software engineering. Our future work involves conducting in-depth interviews and focus groups with recruiters, educators, and software engineers to gain a more comprehensive understanding.

\printbibliography

\end{document}